\documentclass[twocolumn]{aastex62}

\usepackage{tabularx}
\usepackage{threeparttable}
\usepackage[utf8]{inputenc}
\usepackage{amsmath}
\usepackage{comment}
\usepackage{color}

\received{\today}
\revised{\today}
\accepted{\today}

\submitjournal{ApJ}

\begin{document}

%%%%%%%%%%%%%%%%%%%%%%%%%%%%%% title %%%%%%%%%%%%%%%%%%%%%%%%%%%%%%%
\title{Internal structure of molecular gas in a main sequence galaxy with a UV clump at $z = 1.45$}

%%%%%%%%%%%%%%%%%%%%%%%%%%%%% authors %%%%%%%%%%%%%%%%%%%%%%%%%%%%%%
\correspondingauthor{Kaito Ushio}
\email{ushio@kusastro.kyoto-u.ac.jp}

\author{Kaito Ushio}
\affiliation{Department of Astronomy, Kyoto University, Kitashirakawa-Oiwake-Cho, Sakyo-ku, Kyoto 606-8502, Japan}

\author{Kouji Ohta}
\affiliation{Department of Astronomy, Kyoto University, Kitashirakawa-Oiwake-Cho, Sakyo-ku, Kyoto 606-8502, Japan}

\author{Fumiya Maeda}
\affiliation{Department of Astronomy, Kyoto University, Kitashirakawa-Oiwake-Cho, Sakyo-ku, Kyoto 606-8502, Japan}

\author{Bunyo Hatsukade}
\affiliation{Institute of Astronomy, Graduate School of Science, University of Tokyo, 2-21-1 Osawa, Mitaka, Tokyo 181-0015, Japan}

\author{Kiyoto Yabe}
\affiliation{Kavli Institute for the Physics and Mathematics of the Universe (WPI), University of Tokyo, Kashiwa, Chiba 277-8583, Japan}

%%%%%%%%%%%%%%%%%%%%%%%%%%%%%%%%%%%%%%%%%%%%%%%%%%%%%%%%%%%%%%%%%%%%
%%%%%%%%%%%%%%%%%%%%%%%%%%%%% abstract %%%%%%%%%%%%%%%%%%%%%%%%%%%%%
%%%%%%%%%%%%%%%%%%%%%%%%%%%%%%%%%%%%%%%%%%%%%%%%%%%%%%%%%%%%%%%%%%%%
\begin{abstract}
We present results of sub-arcsec ALMA observations of CO(2-1) and CO(5-4) toward a massive main sequence galaxy at $z = 1.45$ in the SXDS/UDS field, aiming at examining the internal distribution and properties of molecular gas in the galaxy.
Our target galaxy consists of the bulge and disk, and has a UV clump in the HST images.
The CO emission lines are clearly detected and the CO(5-4)/CO(2-1) flux ratio ($R_{52}$) is $\sim 1$, similar to that of the Milky Way.
Assuming a metallicity dependent CO-to-$\rm H_{2}$ conversion factor and a CO(2-1)/CO(1-0) flux ratio of 2 (the Milky Way value), the molecular gas mass and the gas mass fraction ($f_{\rm gas} = {\rm molecular \ gas \ mass} / ({\rm molecular \ gas \ mass} + {\rm stellar \ mass})$) are estimated to be $\sim 1.5 \times 10^{11} M_{\odot}$ and $\sim 0.55$, respectively.
We find that $R_{52}$ peak coincides with the position of the UV clump and its value is approximately two times higher than the galactic average.
This result implies high gas density and/or high temperature in the UV clump, which qualitatively agrees with a numerical simulation of a clumpy galaxy.
The CO(2-1) distribution is well represented by a rotating disk model and its half-light radius is $\sim 2.3 \rm \ kpc$.
Compared to the stellar distribution, the molecular gas is more concentrated in the central region of the galaxy.
We also find that $f_{\rm gas}$ decreases from $\sim 0.6$ at the galactic center to $\sim 0.2$ at $3 \times$ half-light radius, indicating that the molecular gas is distributed in more central region of the galaxy than stars and seems to associate with the bulge rather than the stellar disk.
\end{abstract}

%%%%%%%%%%%%%%%%%%%%%%%%%%%%% keywords %%%%%%%%%%%%%%%%%%%%%%%%%%%%%

%%%%%%%%%%%%%%%%%%%%%%%%%%%%%%%%%%%%%%%%%%%%%%%%%%%%%%%%%%%%%%%%%%%%
%%%%%%%%%%%%%%%%%%%%%% section 1: introduction %%%%%%%%%%%%%%%%%%%%%
%%%%%%%%%%%%%%%%%%%%%%%%%%%%%%%%%%%%%%%%%%%%%%%%%%%%%%%%%%%%%%%%%%%%
\section{Introduction} \label{sec:intro}

% figure 01: HST composite image
\begin{figure*}[t]
\begin{center}
\includegraphics[scale = .5, bb = 0 0 811 346]{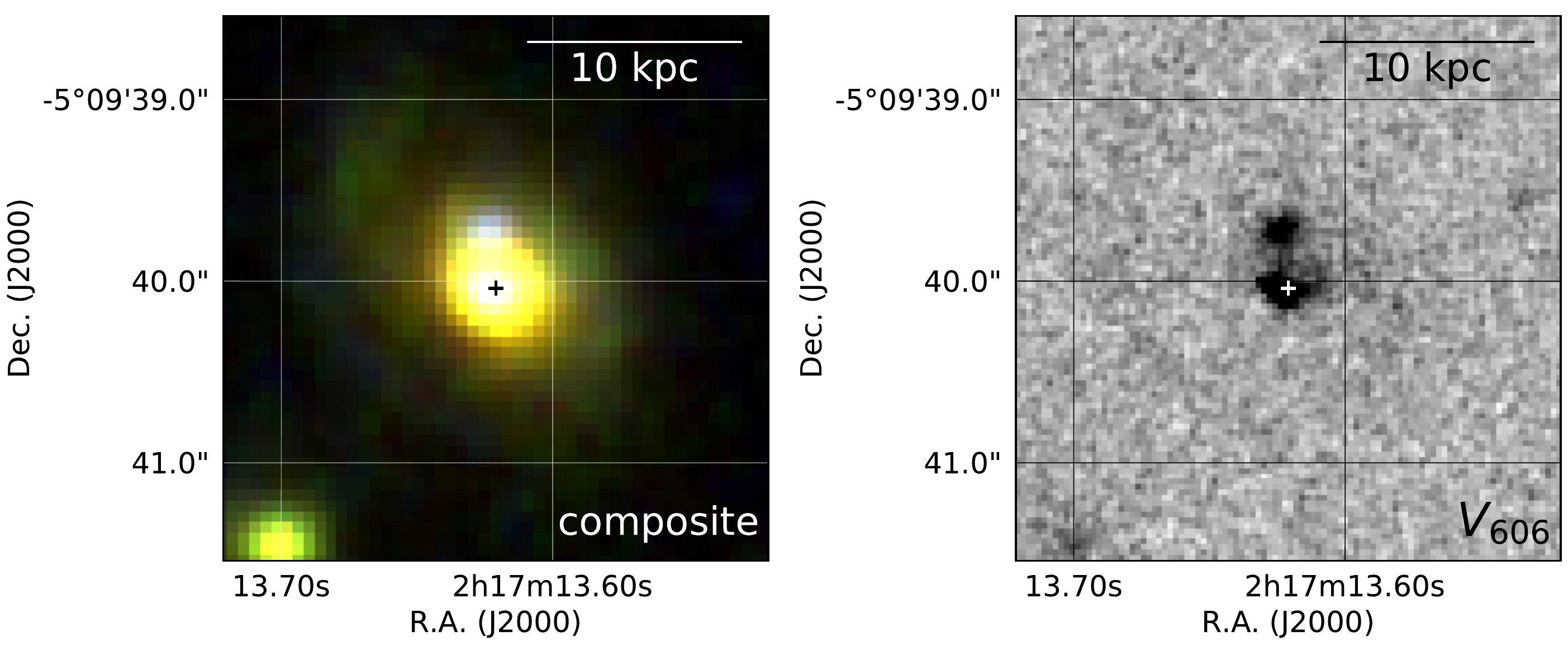}
\end{center}
\caption{\it
  Left: composite image of HST WFC3/IR $H_{160}$, $J_{125}$ and ACS/WFC $V_{606}$ (R, V, U band in the rest-frame, respectively) images of SXDS1\_13015.
  The cross shows the peak position of $J_{125}$.
  %A small object at the bottom left corner is considered not to be relevant to SXDS1\_13015.
  Right: same as the left panel, but for the $V_{606}$ band image.
  SXDS1\_13015 has a UV clump at $\sim 0.\arcsec3$ north of the galactic center (cross mark).
  \label{fig:RGB}
 }
\end{figure*}

Understanding when and how disk galaxies formed is one of the very important subjects in modern astrophysics.
The cosmic star-formation rate (SFR) density increases from the early universe, peaks at $z \sim 1-3$, and decreases by one order of magnitude to the present day \citep[e.g.,][]{Madau14}.
Galaxies are thought to evolve drastically at $z \sim 1-3$, and thus observing galaxies at this redshift range is indispensable to understand galaxy formation and evolution.

Recent deep imagings with high angular resolutions by {\it Hubble Space Telescope} (HST) enable us to resolve the structure of galaxies at $z \sim 1 - 3$ on sub-kiloparsec scale in rest UV to optical bands.
Galaxy population resembling the present-day disk galaxies emerges from $z \sim 2 - 3$ \citep[e.g.,][]{Cameron11, Mortlock13, Talia14} , and at $z \sim 0.8$ fundamental structure of disks seems to be already matured \citep[e.g.,][]{Lilly98, Takeuchi15}.
The bulge-like structures are also seen in disk-like galaxies at $z \sim 1 - 2$ in the rest-frame optical band \citep[e.g.,][]{Cameron11}.
These observational features suggest the disk galaxies acquire their nature seen in the local universe during the period around at $z \sim 1 - 3$.

About half of star-forming galaxies (SFGs) at this redshift range show clumpy structures in rest UV band \citep[e.g.,][]{Elmegreen07,Guo15,Shibuya16}.
These ``clumpy" galaxies are thought to evolve into the present-day disk galaxies.
Observational studies reveal that the clumps have stellar mass of $\sim 10^{8-10} \, M_{\sun}$,  SFR of $\sim 1 - 100 \, M_{\sun}\,\rm yr^{-1}$, size of $1 - 2 \ \rm kpc$, and age of $\sim 10^{7-9} \, \rm yr$.
The clumps close to the center of the host galaxy show larger stellar mass, redder color, and older age than the clumps reside in the outer part \citep[e.g.,][]{Forster11,Guo12,Guo18}.
This trend is explained by a theoretical scenario that a gas-rich disk forms through a cold gas accreting from a dark matter halo and/or streaming from the outside of the galaxy and clumps form through gravitational instability in the gas disk.
Due to the dynamical friction and/or dynamical interaction, the clumps migrate toward the center of the disk and eventually coalesce into a young bulge \citep[e.g.,][]{Noguchi99,Bournaud14}.

Meanwhile the inside-out growth of galaxies is also one aspect of observational results.
For massive SFGs ($M_{\rm star} \gtrsim 10^{11} \, M_{\sun}$), no significant increase of stellar mass is seen in the central region of galaxies since $z \sim 2$, and most of new mass growth is seen in the outer region \citep[e.g.,][]{VanDokkum10,Patel13,Tacchella18}.
For less massive SFGs, the mass growth is seen at all galactic radii, suggesting a synchronous growth of the bulge and disk at $z \ga 1$, while $z \la 1$ in the central region the stellar mass growth gradually halts, despite the mass growth continues in the outer region \citep[e.g.,][]{VanDokkum13,Tacchella18}.
From analysis of main sequence galaxies with $M_{\rm star} > 10^{10} \, M_{\sun}$ at $z \sim 1 - 3$, \citet{Margalef18} show a bulge to disk SFR ratio decreases from $z \sim 3$ to $\sim 1$, which suggests inside-out quenching of star formation.

The properties of molecular gas in SFGs at $z \sim 1 - 3$ have recently been investigated.
The cosmic molecular gas density evolves similarly to the cosmic SFR density: it peaks at $z \sim 1-3$ and is higher than local value by approximately one order of magnitude \citep[e.g.,][]{Maeda17,Decarli19,Riechers19}.
Observations of molecular gas in main sequence SFGs at this redshift range reveal that gas mass fractions ($f_{\rm gas} = M_{\rm gas}/(M_{\rm gas}+M_{\rm star})$, where $M_{\rm gas}$ refers to a molecular gas mass) are high \citep[typically, $f_{\rm gas} \sim 0.5$; e.g.,][]{Tacconi13,Seko16} as compared with those of local galaxies (typically, $f_{\rm gas} \la 0.1$).
The average CO excitation of SFGs (sBzK galaxies) at $z \sim 1.5$ is significantly higher than that of the Milky Way \citep[e.g.,][]{Carilli13,Daddi15}, suggesting the physical conditions of the molecular gas are different.
In order to understand galaxy evolution further, observations of the distribution of molecular gas in normal SFGs including clumpy galaxies on sub-kiloparsec scale are inevitable.
However, resolving distribution of the molecular gas in high-redshift galaxies is still challenging and is limited to some particular cases such as submillimeter galaxies and gravitationally lensed systems \citep[see a review;][]{Hodge20}.
Our knowledge on spatial distribution of molecular gas and its properties in SFGs at $z \sim 1-3$ is still poor.

In this paper, we present results of sub-arcsecond Atacama Large Millimeter/submillimeter Array (ALMA) observations of CO(2-1) and CO(5-4) emission lines toward a massive main sequence galaxy at $z = 1.45$ in the Subaru-XMM/Newton Deep Survey \citep[SXDS;][]{Furusawa08} field.
The CO observations toward this galaxy enable us to study molecular gas properties in a normal star-forming galaxy with a UV clump around at the peak of cosmic SFR density.
Combining multitransition CO data and high-resolution HST data of the galaxy, we also study CO(5-4)/CO(2-1) flux ratio in the UV clump.
Through model fittings, we compare properties of molecular gas and stellar distributions, and also reveal the radial profile of gas mass fraction.

This paper is structured as follows: in Section \ref{sec:sample}, we present the properties of our sample galaxy.
The ALMA observations, the data reductions, and the archival HST data are described in Section \ref{sec:obs}.
The results and discussions are provided in Section \ref{sec:result_and_discussion}.
We report molecular gas properties and distributions in Section \ref{subsec:COresult}, and stellar distributions in Section \ref{subsec:stellardist}.
In Section \ref{subsubsec:distributions}, we compare the molecular gas and stellar distributions.
Finally, we  put the summary in Section \ref{sec:summary}.
Throughout this paper, we adopt a flat cosmology with $\Omega_{M} = 0.3, \ \Omega_{\Lambda} = 0.7$ and $H_0 = 70 \ \rm km \, s^{-1} \, Mpc^{-1}$.
At the redshift of our sample galaxy ($z = 1.450$), the linear scale is $8.45 \ \rm kpc \, arcsec^{-1}$
The initial mass function (IMF) of \citet{Chabrier03} is adopted.
Magnitudes are AB system unless otherwise noted.

%figure 02: 0th moment map and line profile
\begin{figure*}[t]
\begin{center}
\includegraphics[scale = .45, bb =0 0 833 765]{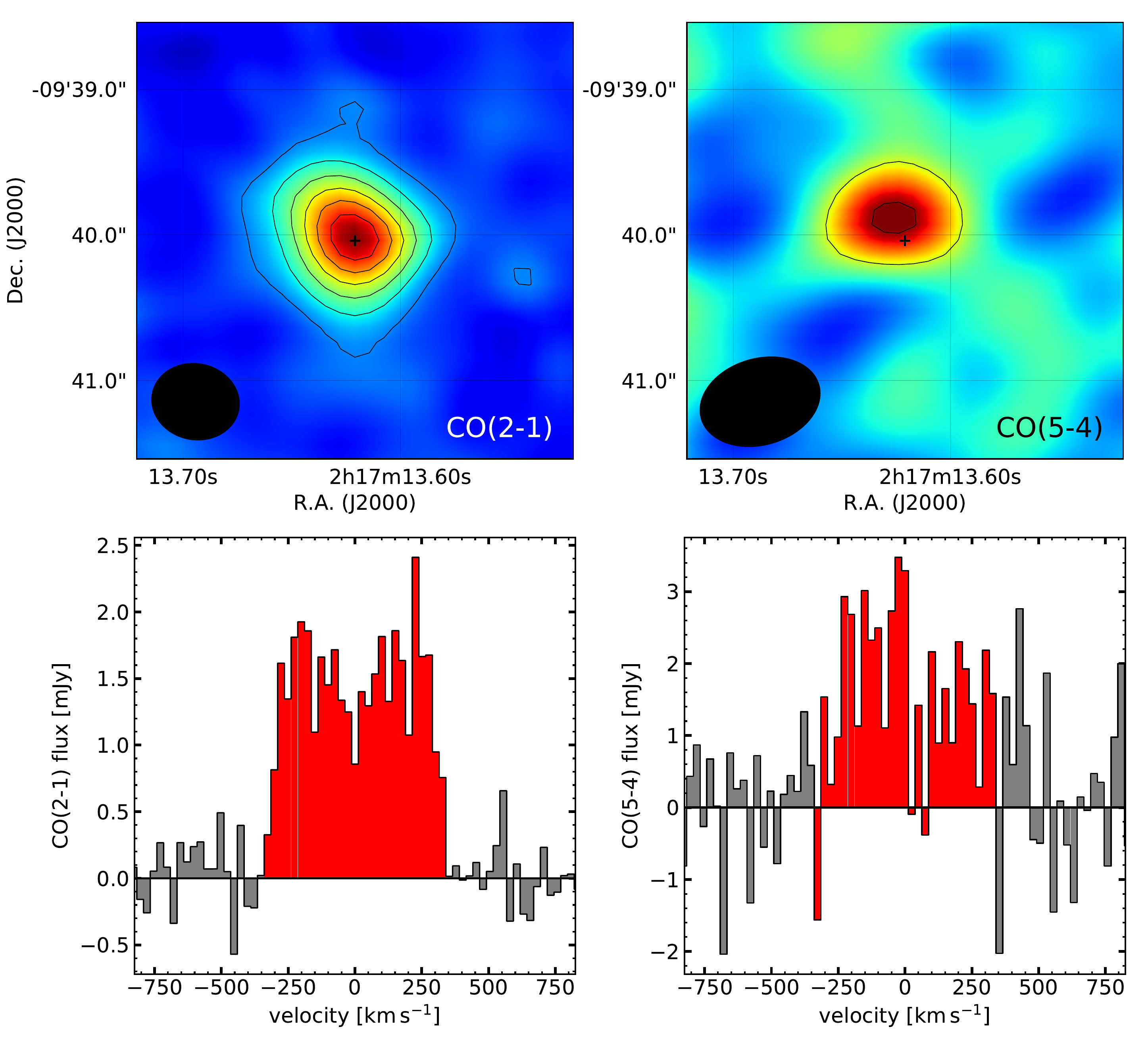}
\end{center}
\caption{\it Top: integrated intensity maps (0th moment maps) of CO(2-1) (left) and CO(5-4) (right) integrated over the velocity range shown in red in the line profiles (bottom panels).
The contours represent $3\sigma,\ 7\sigma,\ 11\sigma,\ \cdots$ ($4 \sigma$ step).
The black crosses show the CO(2-1) peak position.
The filled black ellipse in the bottom left corner shows the synthesized beam size.
Bottom: line profiles of CO(2-1) (left) and CO(5-4) (right) in the 1.\arcsec5 diameter circles.
In the CO(2-1) profile, the emission line of SXDS1\_13015 is shown in red.
In the CO(5-4) profile, the same velocity range with the CO(2-1) profile is also shown in red.}
\label{fig:GalCO}
\end{figure*}

%%%%%%%%%%%%%%%%%%%%%%%%%%%%%%%%%%%%%%%%%%%%%%%%%%%%%%%%%%%%%%%%%%%%
%%%%%%%%%%%%%%%%%%%%%%%%% section 2: sample %%%%%%%%%%%%%%%%%%%%%%%%
%%%%%%%%%%%%%%%%%%%%%%%%%%%%%%%%%%%%%%%%%%%%%%%%%%%%%%%%%%%%%%%%%%%%
\section{Sample Galaxy} \label{sec:sample}

The sample galaxy in this study, SXDS1\_13015, is one of the most massive main sequence galaxies at $z \sim 1.4$ in the sample galaxies by \citet{Yabe12} and \citet{Seko16}.
Using the Fiber Multi-Object Spectrograph (FMOS) on the Subaru telescope, \citet{Yabe12} made near-infrared spectroscopic observations of {\it Ks}-band selected 317 SFGs at $z \sim 1.4$ in the SXDS/UDS \citep[UKIRT Infrared Deep Sky Survey Ultra Deep Survey;][]{Lawrence07} field.
The stellar masses were derived by spectral energy distribution  (SED) fittings with optical to mid-infrared data, adopting the stellar population synthesis model by \citet{Bruzual03} and the IMF by \citet{Salpeter55}.
The SFRs were derived from extinction-corrected rest UV luminosity densities by using a conversion factor by \citet{Kennicutt98annurev}, where the extinction law of \citet{Calzetti00} was assumed and the color excesses were estimated from the rest-frame UV slopes.
After correcting the difference of the adopted IMF with the factor of 0.58 \citep{Speagle14}, the stellar mass and SFR of SXDS1\_13015 are $\sim 1.2 \times 10^{11} \, M_{\sun}$ and $\sim 130 \, M_{\sun}\,\rm yr^{-1}$, respectively.
$\rm H\alpha$ emission lines were detected from 71 galaxies including SXDS1\_13015 and the gas metallicities were derived based on N2 method \citep{Pettini04}.

Using ALMA, \citet{Seko16} made observations of ${\rm ^{12}CO(5-4)}$ and dust thermal emission toward 20 SFGs selected from the sample galaxies by \citet{Yabe12}.
20 SFGs were selected to cover a wide range of stellar mass ($4 \times 10^9 - 4 \times 10^{11} \ M_{\sun}$) and metallicity ($12 + {\rm log(O/H)} = 8.2 - 8.9$) uniformly in the diagrams of stellar mass versus SFR and the stellar mass versus metallicity.
CO(5-4) lines are detected toward 11 galaxies and dust emissions are clearly detected from 2 galaxies.
Both the CO emission and dust thermal emission were detected for SXDS1\_13015.
Adopting metallicity-dependent CO-to-${\rm H_2}$ conversion factor \citep{Genzel12} and CO(5-4)/CO(1-0) luminosity ratio of 0.23 \citep[typical for sBzK galaxies by][]{Daddi15}, the molecular gas mass of SXDS1\_13015 was estimated to be $\sim 1 \times 10^{11} \ M_{\sun}$.
Using a modified blackbody model with dust temperature of $30 \rm \ K$ and dust emissivity index of 1.5, the dust mass of SXDS1\_13015 was estimated to be $\sim 5 \times 10^{8} \ M_{\sun}$.

Properties of SXDS1\_13015 are summarized in Table \ref{table:01}.
Figure \ref{fig:RGB} (left) shows the composite image of HST WFC3 F160W (hereafter, $H_{160}$), F125W ($J_{125}$) and ACS F606W ($V_{606}$).
The galaxy seems to consist of the bulge and disk.
In the disk, an arm-like feature is seen in particular northeastern side of the galaxy.
Figure \ref{fig:RGB} (right) shows HST $V_{606}$ image corresponding to rest-frame UV ($\sim 2500$ \AA).
The image clearly shows the presence of a UV clump located at $\sim 0.\arcsec3$ north of the center, which is also a UV bright region.

%table 01: fundamental properties of SXDS1_13015
\setcounter{table}{0}
\begin{table}[h] % tatenaga
\renewcommand{\thetable}{\arabic{table}}
\centering
\caption{\it Properties of SXDS1\_13015} \label{table:01}
\begin{threeparttable}
\begin{tabularx}{80mm}{>{\centering\arraybackslash}p{40mm}>{\centering\arraybackslash}p{35mm}}
\tablewidth{0pt}
\hline\hline
R.A. (J2000)\tnote{a}            & $\rm 02^{h}17^{m}13.62^{s}$   \\
decl. (J2000)\tnote{a}           & -05\arcdeg09\arcmin40.\arcsec0\\
$z_{\rm CO}$\tnote{b}            & $1.4500 \pm     0.0002$       \\
$M_{\rm star}\,(M_{\sun})$\tnote{c,d} & $1.2^{+0.3}_{-0.2} \times 10^{11}$ \\
SFR ($M_{\sun} \, \rm yr^{-1}$)\tnote{d} & $128 \pm 32$          \\
$12 + \rm log(O/H) $\tnote{e}    & $8.85   \pm     0.04$         \\
$E(B-V) \, (\rm mag)$ \tnote{f}  & 0.49                          \\
%$M_{mol;{\rm CO(5-4)}}\,(M_{\sun})$\tnote{g} & $(1.1 \pm 0.2) \times 10^{11}$\\
%$M_{dust}\,(M_{\sun})$\tnote{g}  & $(4.7 \pm 0.6) \times 10^{8}$\\
\hline\hline
\end{tabularx}
\begin{tablenotes}\footnotesize
  \item[a] {\it Coordinate of the peak pixel of the CO(2-1) 0th moment map.}
  \item[b] {\it Redshift derived from the CO(2-1) line profile.}
  \item[c] {\it Derived from the SED fitting with optical to mid-infrared data.}
  \item[d] {\it Chabrier IMF is adopted. Difference of the adopted IMF is corrected with a factor of 0.58 \citep{Speagle14}. The SFR is derived from the extinction-corrected UV luminosity density.}
  \item[e] {\it Derived with the N2 method. \citet{Pettini04} calibration is adopted.}
  \item[f] {\it Derived from the rest-frame UV slope.}
  %\item[g] {\it Derived from CO(5-4). metallicity-dependent CO-to-${\rm H_2}$ conversion factor \citep{Genzel12} and CO(5-4)/CO(1-0) luminosity ratio of 0.23 \citep{Daddi15} are adopted.}
  %\item[g] {\it The modified blackbody model is adopted.}
\end{tablenotes}
\end{threeparttable}
\end{table}

%%%%%%%%%%%%%%%%%%%%%%%%%%%%%%%%%%%%%%%%%%%%%%%%%%%%%%%%%%%%%%%%%%%%
%%%%%%%%%%%%%%%%%%%%%% section 3: observation %%%%%%%%%%%%%%%%%%%%%%
%%%%%%%%%%%%%%%%%%%%%%%%%%%%%%%%%%%%%%%%%%%%%%%%%%%%%%%%%%%%%%%%%%%%
\section{Data Sources} \label{sec:obs}

%%%%%%%%%%%%%%%%%%%%% subsection 3.1: ALMA data %%%%%%%%%%%%%%%%%%%%
\subsection{CO Data} \label{subsec:COobs}

Observations of ${\rm ^{12}CO(2-1)}$ toward SXDS1\_13015 were made  with ALMA on 2016 August 24 and September 3 during the ALMA Cycle3 (ID: 2015.1.01129.S, PI: K. Ohta).
The number of 12 m antennas was 41.
The length of the longest and shortest baseline was 1.8 km and 15.1 m, respectively, corresponding to an angular resolution of $\sim 0.\arcsec55$ and a maximum recoverable scale of $\sim 11\arcsec$.
The observed frequency range was $93.598 - 94.535 \, \rm GHz$ (band 3) to detect ${\rm ^{12}CO(2-1)}$ emission line ($\nu_{\rm rest} = 230.538 \, \rm GHz$, $\nu_{\rm obs} = 94.097 \, \rm GHz$) with bandwidth of 937.5 MHz and spectral resolution of 564.453 kHz, corresponding to a velocity range and resolution of $2988 \ \rm km \, s^{-1}$ and $1.8 \ \rm km \, s^{-1}$, respectively.
The total on-source time was 4.7 hours.
J0006-0623 and J0238-1636 were used as the flux, bandpass, and amplitude calibrators.
The phase calibrators were J0209-0438 and J0215-0222.

The details of CO(5-4) observations are described by \citet{Seko16}.
The observation was conducted during the ALMA Cycle 0 (ID: 2011.0.00648.S, PI: K. Ohta), and the angular resolution was $\sim$ 0.\arcsec7.
The observed frequency range was $222.094 - 252.583 \, \rm GHz$ (band 6), the spectral resolution was 488.28 kHz ($\sim 0.6 \ \rm km \, s^{-1}$) and the on-source time was $\sim$ 10 minutes.

The raw visibility data of CO(2-1) was calibrated with the Common Astronomy Software Applications \citep[CASA;][]{McMullin07} version 4.7.2 and the observatory-provided calibration script.
The raw visibility data of CO(5-4) was calibrated with CASA version 4.2 by \citet{Seko16}.
The imagings of the ALMA data were carried out using CASA version 5.4.0.
In order to make cleaned channel maps of CO(2-1) and CO(5-4), we used the CASA tasks {\tt uvcontsub} and {\tt tclean}.
Firstly, continuum emission in CO(5-4) data was subtracted with the {\tt uvcontsub}.
After that, we made $3\sigma$-clean channel maps of CO(2-1) and CO(5-4) using the {\tt tclean} with weighting of briggs (${\tt robust} = 0.5$) and channel width of $25 \ \rm km \, s^{-1}$.
The synthesized beam sizes of CO(2-1) and CO(5-4) channel maps were $0.\arcsec61 \times 0.\arcsec52$ ($\sim 4.8 \ \rm kpc$) and $0.\arcsec84 \times 0.\arcsec60$ ($\sim 6.0 \ \rm kpc$), respectively.
The noise levels of CO(2-1) and CO(5-4) at the velocity width of $25 \ \rm km \, s^{-1}$ were $0.18 \, \rm mJy \, beam^{-1}$ and $0.91 \, \rm mJy \, beam^{-1}$, respectively.

%%%%%%%%%%%%%%%%%%%%% subsection 3.2: HST data %%%%%%%%%%%%%%%%%%%%%
\subsection{HST Data} \label{subsec:HSTdata}

SXDS1\_13015 is located in the UKIDSS UDS field, and imaged with HST ACS, WFC3/UVIS and WFC3/IR as a part of Cosmic Assembly Near-infrared Deep Extragalactic Legacy Survey \citep[CANDELS;][]{Grogin11,Koekemoer11} observations.
We used the archival ACS $V_{606}$, WFC3 $J_{125}$ and $H_{160}$ data on CANDELS website\footnote{\url{https://archive.stsci.edu/pub/hlsp/candels/uds/uds-tot/v1.0/}}.
The images are drizzled and the pixel scales are $0.\arcsec03$ for $V_{606}$, and $0.\arcsec06$ for $J_{125}$ and $H_{160}$.
%The ACS F606W, WFC3 F160W and F125W filters are used as rest-frame {\it U}, {\it V} and {\it R} band of the galaxy, respectively.

%%%%%%%%%%%%%%%%%%%%%%%%%%%%%%%%%%%%%%%%%%%%%%%%%%%%%%%%%%%%%%%%%%%%
%%%%%%%%%%%%%%%%%%%%%%%% section 4: results %%%%%%%%%%%%%%%%%%%%%%%%
%%%%%%%%%%%%%%%%%%%%%%%%%%%%%%%%%%%%%%%%%%%%%%%%%%%%%%%%%%%%%%%%%%%%
\section{results and discussions} \label{sec:result_and_discussion}

%%%%%%%%%%%%%%%%%%% subsection 4.1: molecular gas %%%%%%%%%%%%%%%%%%
\subsection{Molecular Gas Properties}\label{subsec:COresult}

The integrated CO(2-1) and CO(5-4) intensity maps (0th moment maps) and the line profiles of SXDS1\_13015 are shown in Figure \ref{fig:GalCO}.
The symmetric double-peak line profile of CO(2-1) implies the galaxy has a rotating gas disk.
The peak position of the CO(5-4) image shows a slight ($\sim 0.\arcsec15$) offset from that of CO(2-1) image and asymmetry is seen in the CO(5-4) line profile.
These features are originated from the presence of the UV clump as described later (Section \ref{subsubsec:excitationmap}).
Thus we adopt the peak in the CO(2-1) map as the center of the galaxy and the central frequency of the observed CO(2-1) emission line ($94.097 \pm 0.008 \ \rm GHz$, corresponding to the redshift of $1.4500 \pm 0.0002$) is taken as the zero velocity.

%%%%%%%%%%%%% subsubsection 4.1.1: molecular gas mass %%%%%%%%%%%%%%
\subsubsection{Total Flux and Molecular Gas Mass}\label{subsubsec:totalmass}

%Table 02: result of CO
\begin{table}[t]
\renewcommand{\thetable}{\arabic{table}}
\centering
\caption{\it Molecular gas properties of SXDS1\_13015} \label{table:molecule}
\begin{threeparttable}
\begin{tabularx}{80mm}{>{\centering\arraybackslash}p{40mm}>{\centering\arraybackslash}p{35mm}}
\tablewidth{0pt}
\hline\hline
$S_{\rm CO(2-1)}\Delta v\,(\rm Jy\,km\,s^{-1})$ & $1.03 \pm 0.05$\\
$S_{\rm CO(5-4)}\Delta v\,(\rm Jy\,km\,s^{-1})$ & $1.16 \pm 0.17$\\
$R_{52}$ \tnote{a}               & $1.1  \pm 0.2$                \\
$L'_{\rm CO(1-0)}\,(\rm K\,km\,s^{-1}\,pc^2)$\tnote{b} & $(5.62 \pm 0.27) \times 10^{10}$\\
$M_{\rm mol}\,(M_{\sun})$\tnote{c} & $(1.45 \pm 0.07) \times 10^{11}$\\
$f_{\rm gas}$ \tnote{d}            & $0.55 \pm 0.08$                 \\
$\tau_{\rm depl} \, (\rm Gyr)$ \tnote{e} & $1.1 \pm 0.3 $            \\
\hline\hline
\end{tabularx}
\begin{tablenotes}\footnotesize
  \item[a] {\it CO(5-4)/CO(2-1) flux ratio.}
  \item[b] {\it We adopt CO(2-1)/CO(1-0) flux ratio of 2 (the Milky Way like value).}
  \item[c] {\it Including a 36\% mass contribution of helium. The metallicity dependent CO-to-$H_2$ conversion factor (Equation (7) of \citet{Genzel15}) is adopted. }
  \item[d] {\it Gas mass fraction ($= M_{\rm mol} / (M_{\rm mol} + M_{\rm star})$).}
  \item[e] {\it Gas depletion time ($= M_{\rm mol} / SFR$)}
\end{tablenotes}
\end{threeparttable}
\end{table}

%Figure 03: CO flux ratio map
\begin{figure}[t]
\begin{center}
\includegraphics[scale = .45, bb = 0 0 524 416]{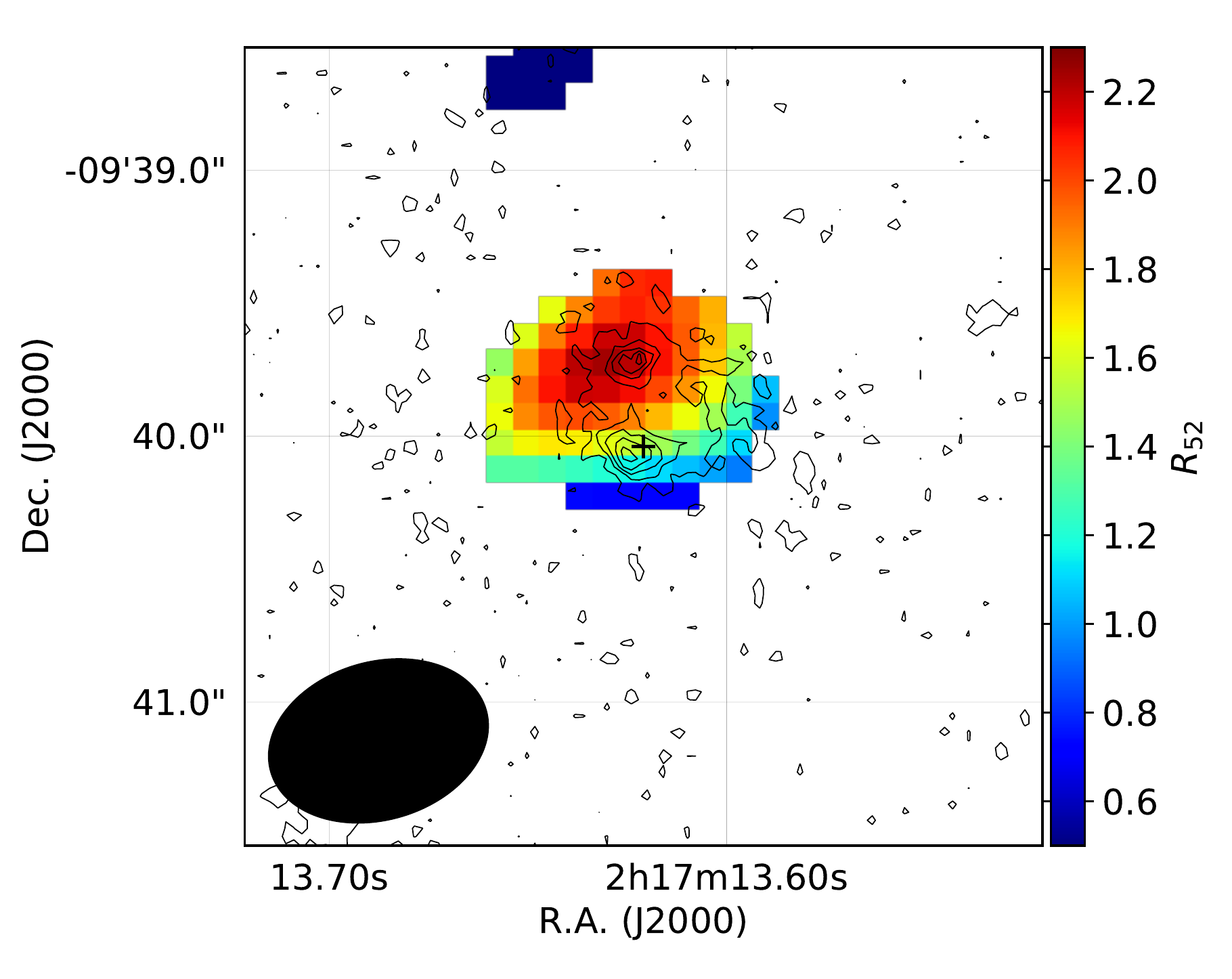}
\end{center}
\caption {\it Spatial distribution of $R_{52}$ in SXDS1\_13015.
The pixels with ${\rm S/N} > 2.5$ in the CO(5-4) map are displayed.
The contours represent $10\%, 30\%, ..., 90\%$ pixel values of the rest UV image relative to the peak of the galaxy.
The black cross and ellipse show the CO(2-1) peak position and the synthesized beam size, respectively.
}
\label{fig:COexcitationMap}
\end{figure}

The total CO(2-1) and CO(5-4) flux of SXDS1\_13015 are derived by fitting an elliptical Gaussian model to the respective integrated intensity maps.
We use the data in a 1.\arcsec5 diameter circle centered on the peak of the CO(2-1) integrated intensity map.
The total CO(2-1) and CO(5-4) flux of SXDS1\_13015 are estimated to be $(1.03 \pm 0.05) \ \rm Jy \, km \, s^{-1}$ and $(1.16 \pm 0.17) \ \rm Jy \, km \, s^{-1}$, respectively.
The beam-deconvolved FWHM of the CO(2-1) source is $4.0 \pm 0.5 \ \rm kpc$.
%The data region used for the flux estimations is broader than the apparent CO(5-4) distribution.
%The fitted flux value, however, does not depend on the shape or size of fitting regions if they include the peak pixel and are not too small.
The CO(5-4) flux value is different from that derived by \citet{Seko16} because of the differences of the adopted velocity width and the data reduction.
%; \citet{Seko16} derived the total CO(5-4) flux using a natural-weighted dirty 0th moment map integrated over the velocity width of $800 \ \rm km \, s^{-1}$.

The CO(5-4)/CO(2-1) flux ratio ($R_{52} = \\ S_{\rm CO(5-4)}\Delta v / S_{\rm CO(2-1)}\Delta v$) is $1.1 \pm 0.2$, which suggests that the CO excitation ladder of SXDS1\_13015 is similar to that of the Milky Way rather than color-selected SFGs at $z \sim 1 - 3$ \citep{Carilli13}.\footnote{As compared with the $R_{52}$-SFR surface density relations recently obtained by \citet{Valentino20} and \citet{Boogaard20}, SXDS1\_13015 shows a lower $R_{52}$ at similar SFR surface density among SFGs at $z \sim 1 - 2$.}
Such low CO excitations are also reported in some SFGs at $z \sim 1-2.5$ \citep[e.g.,][]{Decarli16}.

We calculate the total CO(1-0) luminosity ($L'_{\rm CO(1-0)}$) as
\begin{equation}
\begin{split}
  L'_{\rm CO(1-0)} &= 3.25 \times 10^7\nu_{\rm rest:CO(1-0)}^{-2}D_L^2(1+z)^{-1} \\
  & \quad \times R_{21}^{-1}S_{\rm CO(2-1)}\Delta v ,
\label{eq:COluminosity}
\end{split}
\end{equation}
where $L'_{\rm CO(1-0)}$ is measured in ${\rm K\, km\,s^{-1} \, pc^2}$, $\nu_{\rm rest:CO(1-0)}$ is the rest frequency of the CO(1-0) emission line of 115.271 GHz, $D_L$ is the luminosity distance in Mpc, $R_{21}$ is the CO(2-1)/CO(1-0) flux ratio, and $S_{\rm CO(2-1)}\Delta v$ is the observed CO(2-1) flux in $\rm Jy\ km\,s^{-1}$.
Because the CO excitation ladder of the galaxy is similar to that of the Milky Way, we assume $R_{21} = 2$.
The calculated total CO(1-0) luminosity of SXDS1\_13015 is $(5.62 \pm 0.27) \times 10^{10} \ {\rm K\, km\,s^{-1} \, pc^2}$.

The molecular gas mass is derived from
\begin{equation}
M_{\rm mol} = \alpha_{\rm CO}L'_{\rm CO(1-0)} ,
\label{eq:gasmass}
\end{equation}
where $\alpha_{\rm CO}$ is the CO-to-$\rm H_{2}$ conversion factor in $M_{\sun} \rm (K\, km\,s^{-1}\, pc^2)^{-1}$.
$\alpha_{\rm CO}$ in SFGs at $z \sim 1-2$ is expected to depend on gas metallicity; the value of $\alpha_{\rm CO}$ is larger in galaxies with lower metallicity \citep{Wolfire10, Bolatto13}.
As the dependence of $\alpha_{\rm CO}$ on metallicity,
%we adopt the geometric mean of \citet{Bolatto13} and \citet{Genzel12} \citep[Equation (2) of][]{Tacconi18}:
we adopt the Equation (7) of \citet{Genzel15}:
\begin{equation}
%{\rm log}(\alpha_{\rm CO}) = -1.3 \times (12 + {\rm log(O/H)})_{\rm Denicol\acute{o} 02} + 12 , % Genzel+12: 2.7
\alpha_{\rm CO} = 4.36 \times 10^{-1.27 \times (12 + {\rm log(O/H)} - 8.67)}, % Genzel+15: 2.575
\label{eq:conversionfactor}
\end{equation}
where $\alpha_{\rm CO}$ includes a 36\% mass contribution of helium and $(12 + {\rm log(O/H)})$ is the metallicity based on \citet{Pettini04} calibration.
The $\alpha_{\rm CO}$ of SXDS1\_13015 is calculated to be $2.58\ M_{\sun} \rm (K\, km\,s^{-1}\, pc^2)^{-1}$.
The resulting total molecular gas mass is $(1.45 \pm 0.07) \times 10^{11} \ M_{\sun}$ and the gas mass fraction is $f_{\rm gas} = 0.55 \pm 0.08$.
The gas depletion time ($\tau_{\rm depl} = M_{\rm mol}/{\rm SFR}$) of the galaxy is estimated to be $= 1.1 \pm 0.3 \ \rm Gyr$.
The derived molecular gas properties of SXDS1\_13015 are listed in Table \ref{table:molecule}.

%%%%%%%%%%%%%%%%% subsubsection 4.1.2: CO flux ratio %%%%%%%%%%%%%%%%
\subsubsection{Spatial Distribution of $R_{52}$}\label{subsubsec:excitationmap}

%figure 04: velocity field (1st moment map)
\begin{figure}[t]
\begin{center}
\includegraphics[scale = .45, bb =0 0 539 416]{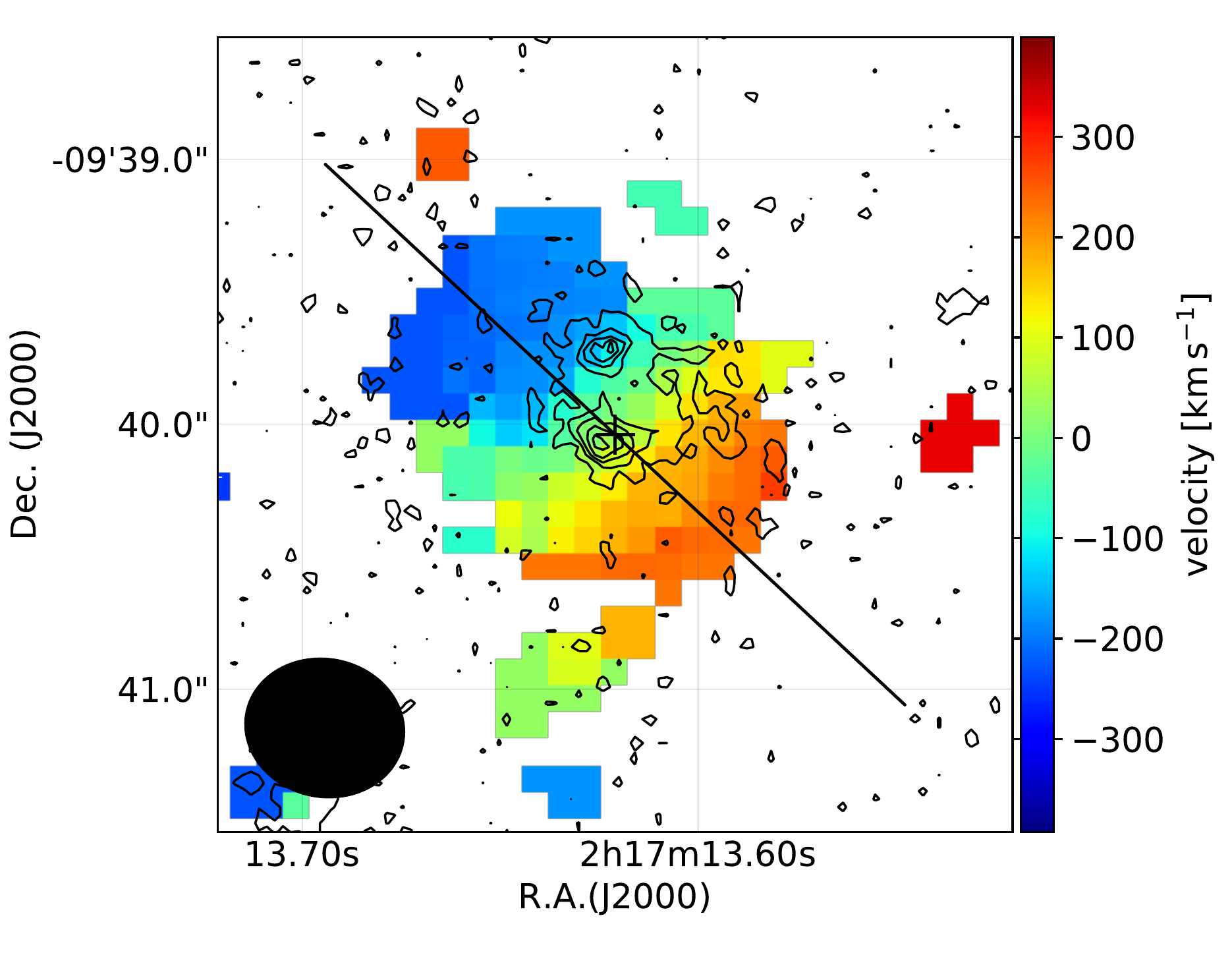}
\end{center}
\caption {\it Velocity field (1st moment map) of SXDS1\_13015.
The contours show the rest UV distribution (same as Figure \ref{fig:COexcitationMap}).
The black cross and solid line show the CO(2-1) peak position and the major axis of the galaxy, respectively.
The filled black ellipse in the bottom left corner shows the synthesized beam size.
}
\label{fig:Galvel}
\end{figure}

%Figure 05: CO(5-4) 0th moment map integrated over clump velocity range
\begin{figure*}[t]
\begin{center}
\includegraphics[scale = .4, bb = 0 0 1048 833]{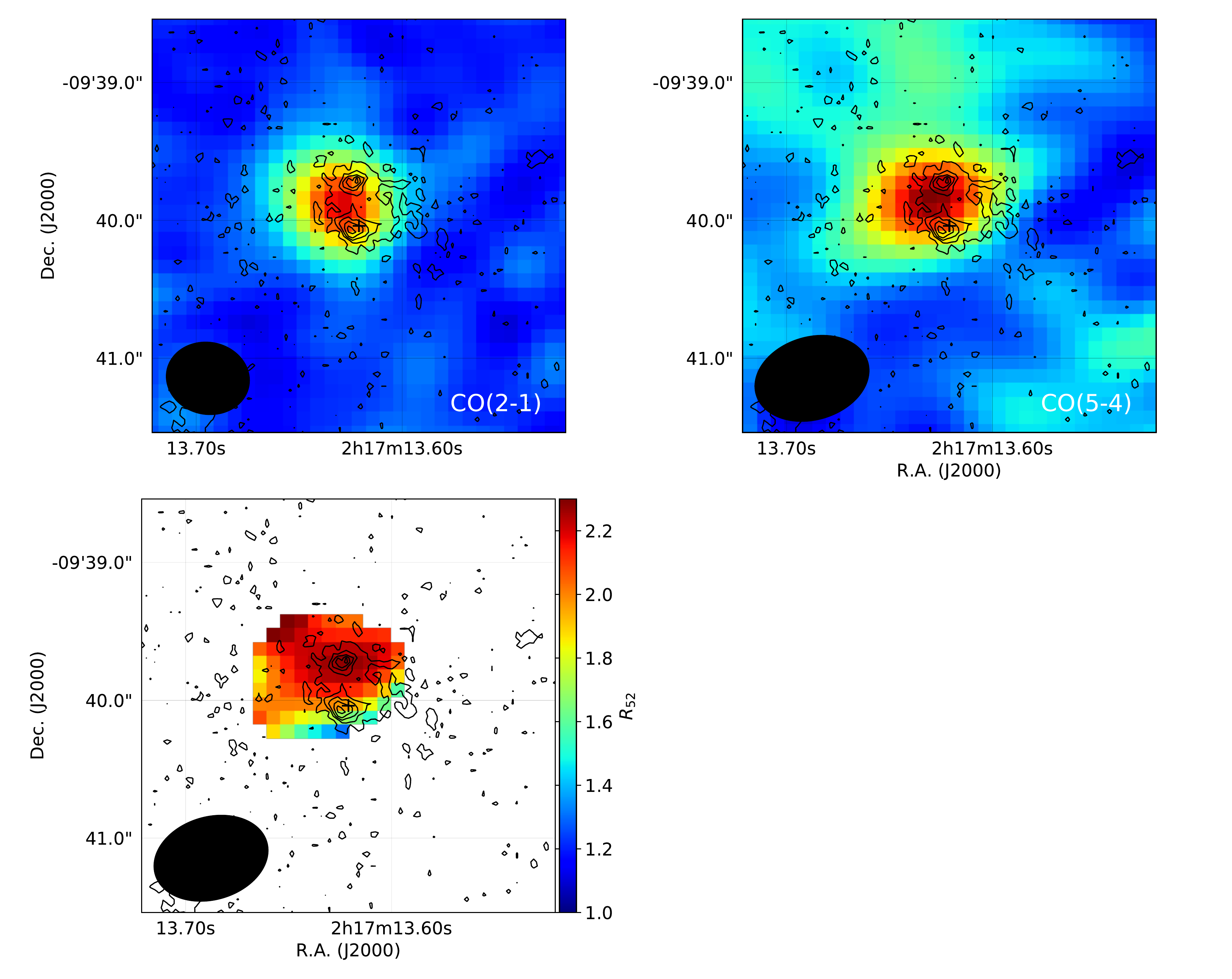}
\end{center}
\caption {\it Top left: CO(2-1) 0th moment map integrated over the velocity range of $-250 \ \rm km \, s^{-1}$ to $0 \ \rm km \, s^{-1}$.
The contours show the rest UV distribution (same as Figure \ref{fig:COexcitationMap}).
The black cross and the ellipse show the CO(2-1) peak position in the total integrated map and the synthesized beam size, respectively.
Top right: same as top left panel, but for CO(5-4).
Bottom left: same as Figure \ref{fig:COexcitationMap}, but for the velocity range of $-250 \ \rm km \, s^{-1}$ to $0 \ \rm km \, s^{-1}$.
}
\label{fig:negvel_mom0}
\end{figure*}

%figure 06: result of GalPaK fitting
\begin{figure*}[t]
\begin{center}
\includegraphics[scale = .4, bb =0 0 906 833]{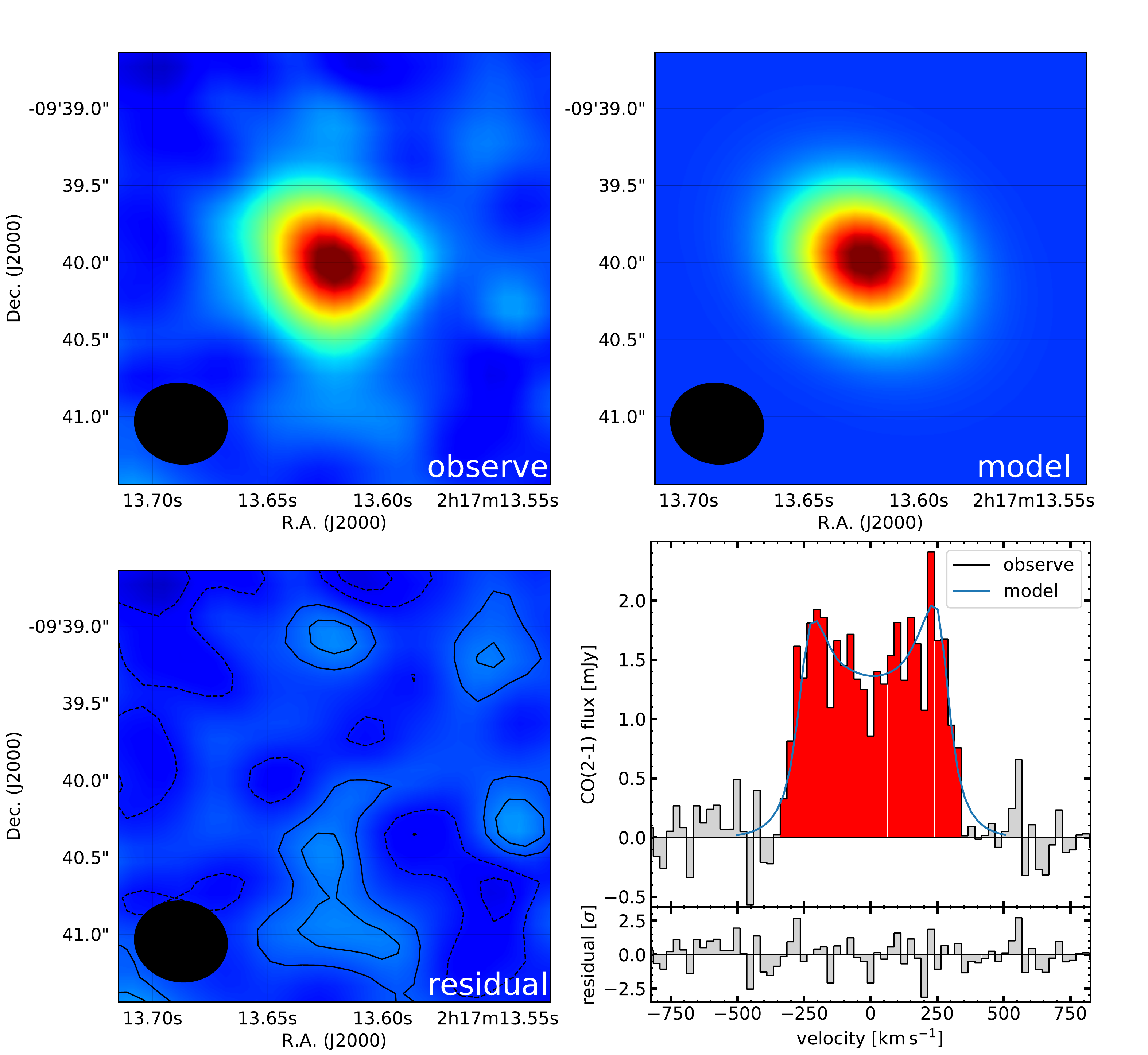}
\end{center}
\caption {\it Top left: observed integrated intensity map of CO(2-1).
The filled black ellipse in the bottom left corner shows the synthesized beam size.
Top right and bottom left: Same as top left, but for best-fit model through ${\rm GalPaK^{3D}}$ and residual ($observed - model$), respectively.
The contours in the residual map represent $-2\sigma,\ -\sigma, \sigma$ and $2\sigma$.
Bottom right: observed and modeled line profiles (red and blue solid curve, respectively).
The residual normalized with $1\sigma$ noise level is shown in lower part.
}
\label{fig:Galpak}
\end{figure*}

We made the $R_{52}$ map by dividing the CO(5-4) 0th moment map by the CO(2-1) map convolved to the same beam size as CO(5-4) data.
The $R_{52}$ is calculated in pixels with ${\rm S/N} > 2.5$ in the CO(5-4) map.
The resulting CO flux ratio map is shown in Figure \ref{fig:COexcitationMap}.
The peak of the CO(5-4)/CO(2-1) flux ratio ($R_{52} \sim 2.2$) is significantly larger than the $R_{52}$ averaged over SXDS1\_13015 and that of the Milky Way.
The ratio is comparable to the average $R_{52}$ for sBzK galaxies at $z \sim 1.5$ \citep{Daddi15}.
The peak of the flux ratio is located at $\sim 0.\arcsec3$ north of the peak of the CO(2-1) distribution (cross mark in Figure \ref{fig:COexcitationMap}).

Interestingly, the location coincides with the UV clump.
The velocity at this position is $\sim -150 \ \rm km \, s^{-1}$ in the CO(2-1) velocity field (Figure \ref{fig:Galvel}), and this velocity corresponds to the peak velocity in the CO(5-4) profile (Figure \ref{fig:GalCO}, bottom right); the asymmetric feature of the CO(5-4) profile is presumably due to this component.
We show the 0th moment maps integrated over the velocity range of $-250 \ \rm km \, s^{-1}$ to $0 \ \rm km \, s^{-1}$ in Figure \ref{fig:negvel_mom0}.
The $R_{52}$ map in this velocity range also peaks at the position of the UV clump.
In the UV clump, it is expected that the density and/or temperature of the molecular gas are higher than those in other regions.
This result, the high $R_{52}$ in the UV clump, is qualitatively consistent with the CO excitation ladder modeled by a large velocity gradient analysis of the hydrodynamically simulated high-$z$ clumpy galaxy \citep{Bournaud15}.

It is worth noting that the spatial offset and the asymmetric profile of CO(5-4) are not results of noise.
We performed simulations of CO(5-4) observation with CASA task {\tt simobserve} assuming a gaussian distribution mimicking the CO(5-4) distribution and line profile.
Consequently, we found that the 1-$\sigma$ uncertainty of the peak position of CO(5-4) is $\sim 0.04 \ \rm arcsec$.
he observed offset between the peak positions of CO(5-4) and CO(2-1) is $\sim 0.15 \ \rm arcsec$ which corresponds to $\sim 3.8 \sigma$.
We also found that the 1-$\sigma$ uncertainty of the peak velocity is $\sim 30 \ \rm km \, s^{-1}$.
The observed peak CO(5-4) velocity and the central CO(2-1) velocity is $\sim 150 \ \rm km \, s^{-1}$ which corresponds to $\sim 5 \sigma$.
Thus the offset and the peak velocity difference are considered to be real.
It should be also noted here, we made astrometric corrections to the HST images with the offset shown below.
We assume the peak of $J_{125}$ image coincides with the peak of the CO(2-1) 0th moment map, because the CO(2-1) shows a good symmetry in the line profile and S/N of the image is high.
The offset is $(\Delta{\rm R.A.}, \Delta{\rm Dec.}) = (+0.\arcsec058, -0.\arcsec198)$.
Similar amounts of the offset between ALMA astrometry and HST astrometry are also reported in Hubble Ultra-Deep Field and CANDELS GOODS-S region \citep{Barro16,Rujopakarn16,Dunlop17}.

\citet{Cibinel17} presented ALMA observations of CO(5-4) line towards a main sequence, clumpy galaxy at $z \sim 1.5$, but no CO(5-4) emission from the UV clumps was detected.
This is presumably due to the low SFR of the UV clumps ($\sim 1 - 4 \ M_{\odot} \rm \, yr^{-1}$).
The sensitivity of their observation was not deep enough to detect the CO(5-4) emissions from the UV clumps.
The SFR of the UV clump in our sample galaxy is estimated to be $\sim 30 \  M_{\odot} \rm \, yr^{-1}$ from UV luminosity, and is larger than those of clumps in \citet{Cibinel17}.
By assuming a depletion time of $0.5 \ \rm Gyr$ and the sBzK-like CO $J$-ladder, the contribution of the UV clump to the total CO(5-4) flux is expected to be $ \sim 20-25 \%$ in our case.

%table 03: best-fit parameters GalPaK
\begin{table}[t]
\renewcommand{\thetable}{\arabic{table}}
\centering
\caption{\it Best-fit parameters derived by ${\rm GalPaK^{3D}}$} \label{table:02}
\begin{threeparttable}
\begin{tabularx}{80mm}{>{\centering\arraybackslash}p{40mm}>{\centering\arraybackslash}p{30mm}}
\tablewidth{0pt}
\hline\hline
flux $\rm (Jy \, km \, s^{-1})$        & $1.10 \pm 0.05$   \\
$r_{1/2}$ (kpc)\tnote{a}               & $2.30 \pm 0.12$   \\
inclination (deg)                      & $47.6 \pm 6.5$    \\
PA (deg)                               & $42.8 \pm 2.0$    \\
$r_{t}$ (kpc)\tnote{b}                 & $0.48 \pm 0.22$   \\
$V_{\rm max}\ \rm(km\,s^{-1})$\tnote{c}& $350  \pm 30$     \\
\hline\hline
\end{tabularx}
\begin{tablenotes}\footnotesize
  \item[] {\it Uncertainties are estimated based on simulations (see text).}
  \item[a] {\it Half-light radius.}
  \item[b] {\it Turnover radius.}
  \item[c] {\it Inclination-corrected maximum rotational velocity.}
\end{tablenotes}
\end{threeparttable}
\end{table}

%Table 04: result of GALFIT
\begin{table*}[t]
\renewcommand{\thetable}{\arabic{table}}
\centering
\caption{\it Results of fitting S\'{e}rsic profile models to $J_{125}$ and $H_{160}$ images} \label{table:03}
\begin{threeparttable}
\begin{tabularx}{130mm}{>{\centering\arraybackslash}p{45mm}>{\centering\arraybackslash}p{40mm}>{\centering\arraybackslash}p{40mm}}
\tablewidth{0pt}
\hline\hline
filter                       & $J_{125}$     & $H_{160}$ \\
\hline
$n$ \tnote{a}                & $2.04 \pm 0.19$ & $1.76 \pm 0.11$ \\
$R_{1/2}$ (kpc) \tnote{a}    & $3.54 \pm 0.15$ & $3.22 \pm 0.07$ \\
$b/a$ \tnote{a}              & $0.62 \pm 0.01$ & $0.60 \pm 0.01$ \\
PA (deg) \tnote{a}           & $32.0 \pm 0.9$  & $36.2 \pm 0.7$  \\
$R_{1/2; \rm bulge}$ (kpc) \tnote{b} & $0.45\pm0.23$ & $0.44\pm0.22$ \\
$R_{1/2; \rm disk}$ (kpc) \tnote{b}  & $5.05\pm0.33$ & $4.49\pm0.21$ \\
$L_{\rm bulge}/L_{\rm disk}$ \tnote{b}  & $0.38^{+0.11}_{-0.08}$ & $0.36^{+0.11}_{-0.08}$ \\
\hline\hline
\end{tabularx}
\begin{tablenotes}\footnotesize
  \item  {\it Uncertainties are estimated by changing sky level within its error, stars to construct the PSFs and the size of the fitting region.}
  \item[a] {\it Results of single component fitting. S\'{e}rsic index, half-light radius, axis ratio, and position angle, respectively.}
  \item[b] {\it Results of two components (bulge ($n=4$) and disk ($n=1$)) fitting.}
\end{tablenotes}
\end{threeparttable}
\end{table*}

%Figure 07: stellar radial profile
\begin{figure*}[t]
\begin{center}
\includegraphics[scale = .25, bb = 0 0 1625 794]{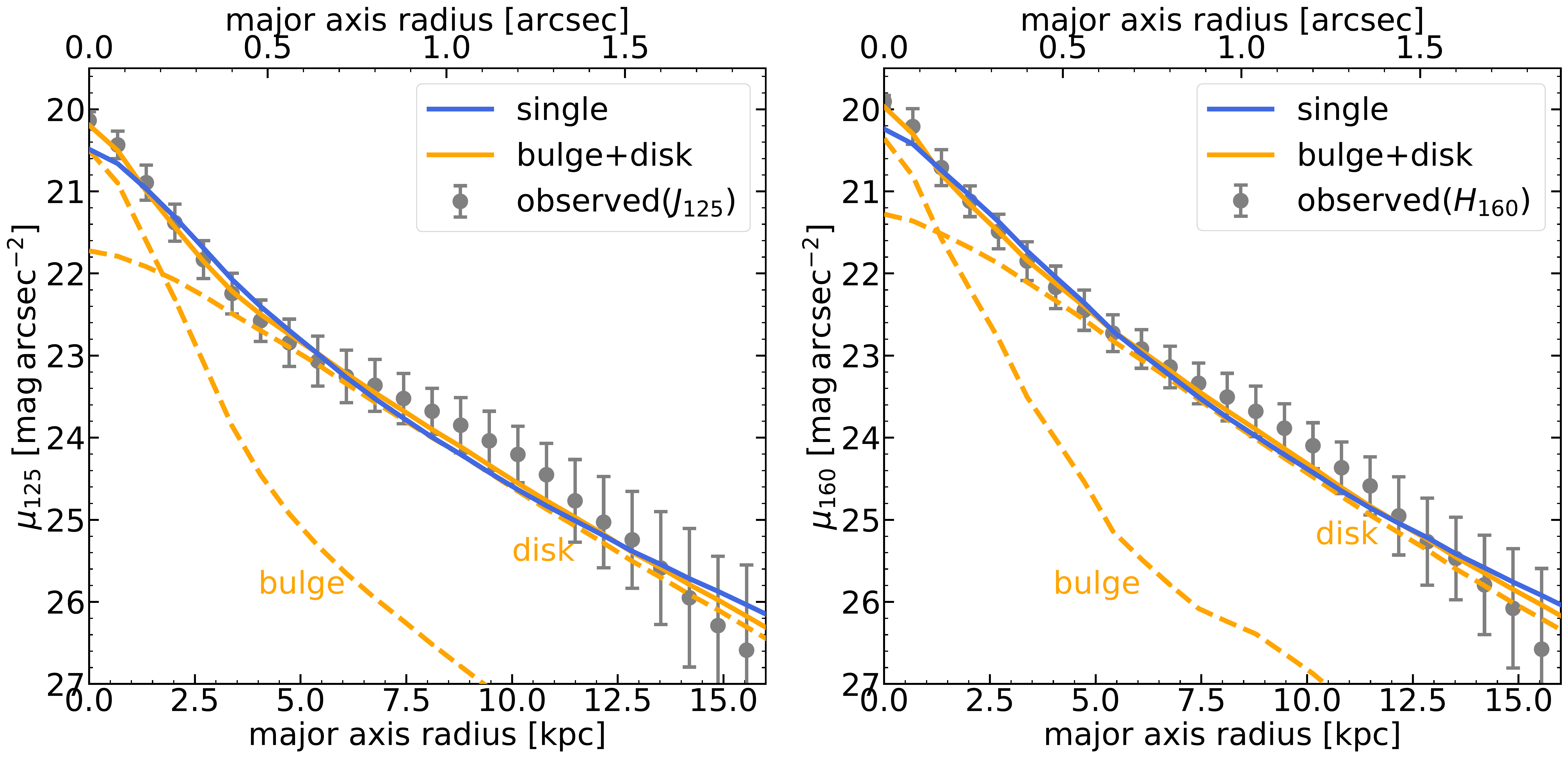}
\end{center}
\caption {\it Left: $J_{125}$ radial profiles of SXDS1\_13015.
$\mu_{125}$ is the surface brightness.
Gray circles with errorbars show the observed radial profile.
The best-fit single component model and two components (bulge + disk) model are shown with blue and orange solid line, respectively.
Dashed orange lines show the bulge and disk components of the best-fit two components model.
Right: same as the left panel, but for $H_{160}$.
}
\label{fig:radialprofile}
\end{figure*}

%%%%%%%%%%%%%%%%%%% subsubsection 4.1.4: GalPaK %%%%%%%%%%%%%%%%%%%%
\subsubsection{Fitting a Rotating Disk Model}\label{subsubsec:rotation}

The velocity field of CO(2-1) (Figure \ref{fig:Galvel}) shows a clear velocity gradient along the major axis, indicating that the galaxy has a rotating gas disk, which is also implied by the double-peak CO(2-1) line profile.
In order to eliminate the effect of beam smearing and derive the intrinsic {molecular gas distribution and} kinematic properties of the galaxy, we fit a rotating disk model by using ${\rm GalPaK^{3D}}$ \citep[Galaxy Parameters and Kinematics;][]{Bouche15}.
${\rm GalPaK^{3D}}$ is a program code that estimates morphological parameters (e.g., size and inclination) and kinematic parameters (e.g., maximum rotational velocity and velocity dispersion) of the galaxy by fitting a disk model convolved with two-dimensional PSF and one-dimensional line spread function (LSF) to a given data cube.
We fit a disk model whose radial flux profile is exponential (S\'{e}rsic index $n=1$) and radial rotational velocity profile is hyperbolic tangent to the CO(2-1) channel map.
PSF and LSF correspond to the synthesized beam ($0.\arcsec61 \times 0.\arcsec53$, ${\rm PA} = 79\degr$) and the channel bin (velocity width is $25 \ \rm km \, s^{-1}$), respectively.
The fitting parameters are central position, central frequency, total flux, half-light radius ($r_{1/2}$), inclination, position angle, turnover radius ($r_{t}$), maximum rotational velocity ($V_{\rm max}$), and velocity dispersion ($\sigma_0$) which is assumed to be constant over the disk.
The best-fit models as well as input image are shown in Figure \ref{fig:Galpak} (upper panels).
The residual ($\rm observed - model$) integrated intensity map and line profile are also shown in Figure \ref{fig:Galpak} (bottom panels).
The residuals demonstrate that the CO(2-1) distribution of the galaxy is well represented by the rotating disk model.
The best-fit parameters are listed in Table \ref{table:02}.
In order to estimate the uncertainties on the best-fit parameters, we created 100 cleaned cubes by simulating observations of the best-fit disk model with {\tt simobserve} by changing the noise seeds randomly.
Disk models are fitted to the simulated cubes by ${\rm GalPaK^{3D}}$, and the means and the standard deviations of the disk model parameters are derived.
We adopt the square root of the sum of the square of systematic error and standard deviation as the uncertainty for each best-fit parameter derived by ${\rm GalPaK^{3D}}$.
The best-fit flux ($1.10 \pm 0.05 \ \rm Jy \, km \, s^{-1}$) is consistent with that derived from the two-dimensional Gaussian fitting to the integrated intensity map.
The best-fit $r_{1/2} = 2.30 \pm 0.12 \ \rm kpc$ is also consistent with the HWHM of the 2D Gaussian fitting ($\sim 2 \ \rm kpc$)\footnote{Generally, the HWHM of a 2D Gaussian corresponds to half-light radius.}.
The best-fit $V_{\rm max}$ and $\sigma_0$ are $350 \pm 30 \ \rm km \, s^{-1}$ and $2.7 \pm 8.7 \ \rm km \, s^{-1}$, respectively.
The best-fit value for $\sigma_0$ is very much small and has a very large uncertainty ($\sigma_0$ is derived to be $\sim 40 \ \rm km \, s^{-1}$ from some simulated cubes).
The value is considered not to be reliable.
In fact, the observed line profile is well reproduced without changing other output parameters such as $r_{1/2}$, even if we fit a disk model with a fixed $\sigma_{0} \sim 50 \ \rm km \, s^{-1}$ \citep[typical value found in $\rm H_{\alpha}$ observations for $z < 1$ disks; e.g.,][]{Wisnioski15,Forster18}.
Due to the large beam size, the observed line width seems to be largely affected by beam smearing effect, and it is hard to constrain the intrinsic velocity dispersion well.

%%%%%%%%%%%%%%% subsection 4.2: stellar distribution %%%%%%%%%%%%%%%
\subsection{Stellar Distribution} \label{subsec:stellardist}

In order to derive properties of the stellar distribution, we fit S\'{e}rsic profile models to HST WFC3/IR $J_{125}$ and $H_{160}$ band images using GALFIT \citep{Peng02}.
GALFIT is a program code that fits given surface brightness model convolved with PSF to an observed galaxy.
Four(Six) bright but unsaturated stars in the $J_{125}$($H_{160}$) image of the CANDELS-UDS field are stacked to construct the PSFs for the fittings.
The $\sim 2.\arcsec5 \times 2.\arcsec5$ boxy region centered at the image peak is used for the fitting.
The pixels contaminated by neighboring objects are masked out.
After sky subtractions, we firstly fit the surface brightness with single component S\'{e}rsic profile.
The fitting parameters are central position, total magnitude, half-light radius ($R_{1/2}$), S\'{e}rsic index ($n$), axial ratio, and position angle.
We also fit another surface brightness model with two-component S\'{e}rsic profiles.
In this fitting, S\'{e}rsic indices are fixed to be $n=4$ for bulge and $1$ for disk, because the galaxy has a rotating gas disk and seems to consist of the bulge and disk in the HST composite image (see Figure \ref{fig:RGB}).
The fitting parameters are total magnitude of each component, half-light radius of each component ($R_{1/2; \rm bulge}, R_{1/2; \rm disk}$) and axial ratio of the bulge.
Central position of each component, position angle of each component, and axial ratio of the disk are fixed to the values derived from the single component fitting.
The derived best-fit parameters are listed in Table \ref{table:03}.
The observed and fitted radial profiles are shown in Figure \ref{fig:radialprofile}.

%Figure 08: comparison with Patel+13
\begin{figure}[t]
\begin{center}
\includegraphics[scale = .5, bb = 0 0 433 398]{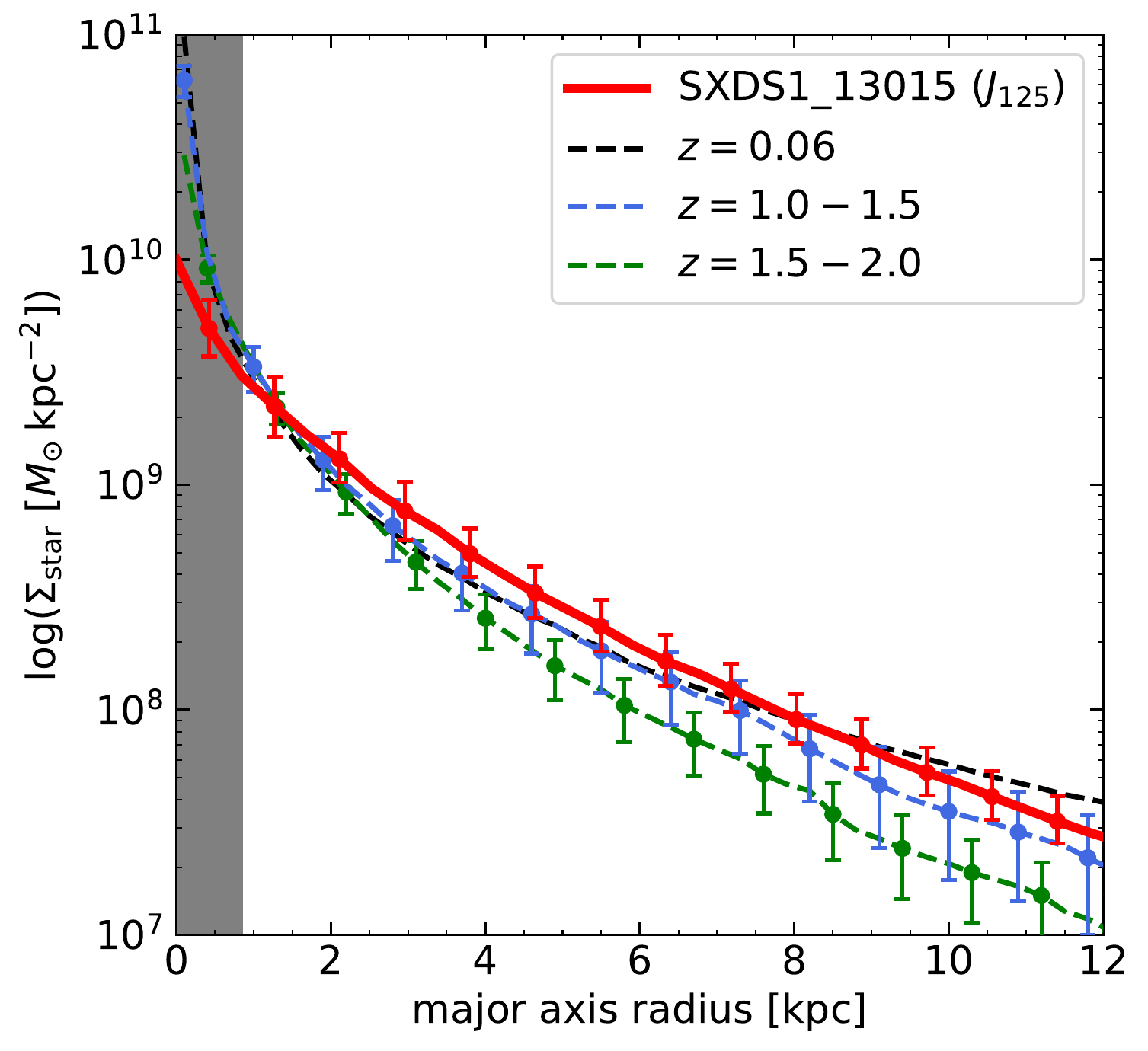}
\end{center}
\caption {\it The red line shows the stellar mass surface density profile of SXDS1\_13015 with the uncertainty derived from the error of the stellar mass and galaxy parameters of GALFIT.
A constant mass-to-luminosity ratio is assumed.
The green, blue, and black dashed lines show the median profiles of galaxies at $z = 1.5 - 2.0$, $1.0 - 1.5$, and $0.06$, respectively \citep{Patel13}.
The galaxies are selected at a constant cumulative number density of $1.4 \times 10^{-4} \ \rm Mpc^{-3}$ for different redshifts (corresponding to the stellar mass of $\sim 1.1 \times 10^{11} \ M_{\sun}$ for $1.0 < z < 1.5$).
We show the standard deviation of each median profile (see text).
The gray shaded region shows the HWHM of PSFs for our fitting.
}
\label{fig:comparePatel}
\end{figure}

%Figure 09: intrinsic distributions
\begin{figure*}[t]
\begin{center}
\includegraphics[scale = .25, bb = 0 0 1900 768]{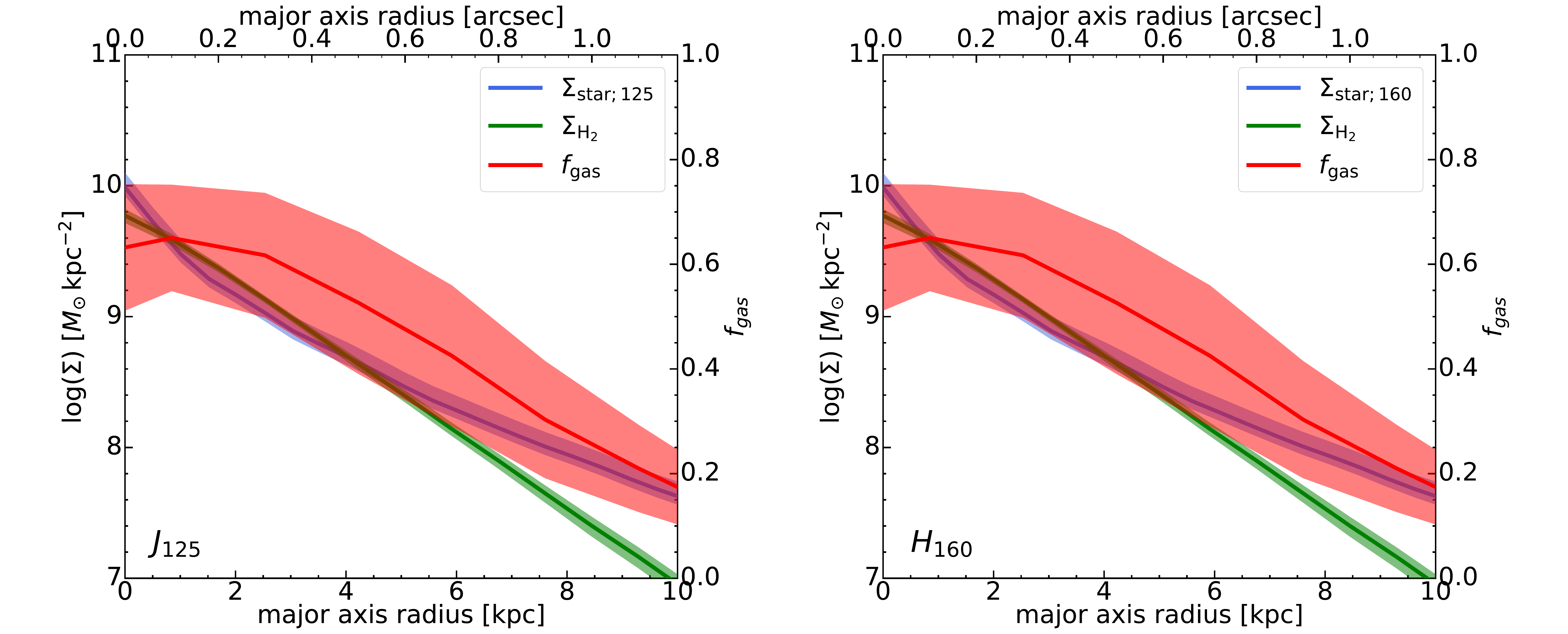}
\end{center}
\caption {\it Left: Modeled (intrinsic) molecular gas and stellar ($J_{125}$) radial distributions of SXDS1\_13015.
Green and blue line shows the molecular gas and stellar (one component) surface density profile, respectively.
Shaded regions show the uncertainties.
The gas mass fraction ($f_{\rm gas} = M_{\rm mol}/(M_{\rm mol}+M_{\rm star})$) is also plotted with red line, and the shaded region shows the uncertainty of $f_{\rm gas}$.
Right: same as the left panel, but for $H_{160}$.
}
\label{fig:intrinsicprofile}
\end{figure*}

Assuming a constant mass-to-luminosity ratio, we derived the stellar mass surface density profile of SXDS1\_13015 from $J_{125}$ distribution modeled with the single component model.
In Figure \ref{fig:comparePatel}, the derived profile of SXDS1\_13015 is plotted with the median profiles of galaxies at different redshifts by \citet{Patel13}.
The sample galaxies of \citet{Patel13} are selected at a constant cumulative number density of $1.4 \times 10^{-4} \ \rm Mpc^{-3}$ for different redshifts (corresponding to the stellar mass of $\sim 1.1 \times 10^{11} \ M_{\sun}$ for $1.0 < z < 1.5$).
The number of sample galaxies at $1.0 < z < 1.5$ is 20 including $\sim 65\%$ of quiescent galaxies.
The surface density profiles are derived by assuming a constant mass-to-luminosity ratio, and are modeled by fitting single component S\'{e}rsic profile to the rest optical image for each redshift using GALFIT.
The standard deviations for the median profiles at $z \sim 1.0-1.5$ and $z \sim 1.5-2.0$ are calculated by generating 1000 median profiles based on the uncertainties of the median half-light radius and S\'{e}rsic index at each redshift provided by \citet{Patel13}.
SXDS1\_13015 shows slightly higher surface density than the median of the galaxies at similar redshift, and reaches to the surface density comparable to that of galaxies at $z \sim 0.06$.
The surface density of SXDS1\_13015 is also comparable to that of local galaxies with the stellar mass of the Milky Way \citep{VanDokkum13}.
At $r \la 1 \ \rm kpc$, the surface density of SXDS1\_13015 is lower than those of the median profiles.
However, the scale of $1 \ \rm kpc$ is comparable to the HWHM of the PSF for our fitting ($\sim 0.8 \ \rm kpc \sim 1.5 \ \rm pixels$), and it is possible that the profile is not sufficiently recovered within this scale.

%%%%%%%%%%%%%% subsection 4.3: intrinsic distribution %%%%%%%%%%%%%%
\subsection{Molecular Gas and Stellar Distribution in SXDS1\_13015} \label{subsubsec:distributions}

%%%%%%%%%%%%% subsubsection 4.3.1: modeled distribution %%%%%%%%%%%%%
\subsubsection{Modeled Gas and Stellar Distributions}
\label{subsubsec:modeldistributions}

By using the modeled molecular gas distribution derived by $\rm GalPaK^{3D}$ and the stellar distribution derived by GALFIT, we obtain intrinsic radial surface distributions of the molecular gas as well as the stellar component in SXDS1\_13015 (Figure \ref{fig:intrinsicprofile}).
The uncertainty on the molecular gas surface distribution is estimated with radial profiles derived from $\rm GalPaK^{3D}$ fitting to the 100 simulated cubes (see Section \ref{subsubsec:rotation}).
The uncertainty on the stellar surface distribution is derived from uncertainties on the total stellar mass and galaxy parameters of GALFIT.
We also derive gas mass fraction ($f_{\rm gas}$) at each radius.
The uncertainty on $f_{\rm gas}$ is derived from the uncertainties on the molecular gas and stellar radial surface distributions.
$f_{\rm gas}$ is higher than galactic total value ($\sim 0.55$) at $r \lesssim 3 \ \rm kpc$, and decreases with increasing galactocentric radius from $\sim 0.6$ at the central region to $\sim 0.2$ at $\sim 3 \times r_{1/2}$.
The molecular gas is distributed in more inner region of the galaxy than stars and seems to associate with the bulge rather than the stellar disk, which is also indicated by the best-fit half-light radii.

By using the best-fit kinematic parameters of the molecular gas disk, dynamical masses ($M_{\rm dyn}(r) \sim rv_{\rm rot}^{2}/G$) at $r = r_{1/2}, \ 2r_{1/2}, \ 3r_{1/2}$ ($r_{1/2}$ is half-light radius of molecular gas disk) are estimated to be $\sim (0.7,\ 1.3,\ 2.0) \times 10^{11} \ M_{\sun}$.
We also derive the molecular gas and stellar masses within these radii assuming the inner mass ($m(<r)$) is proportional to the luminosity within the radius ($L(<r)$) in $J_{125}$ image:
\begin{equation}
m(<r) = \frac{L(<r)}{L_{\rm total}} M_{\rm total} \ .
\label{eq:innermass}
\end{equation}
The molecular gas and stellar masses within $r = r_{1/2}, \ 2r_{1/2}, \ 3r_{1/2}$ are $m_{\rm mol} \sim (0.7,\ 1.2,\ 1.4) \times 10^{11} \ M_{\sun}$ and $m_{\rm star} \sim (0.4,\ 0.7,\ 0.9) \times 10^{11} \ M_{\sun}$, respectively.
Baryon fractions ($f_{\rm baryon} = (m_{\rm mol} + m_{\rm star})/(M_{\rm dyn} + m_{\rm mol} + m_{\rm star})$) within those radii are estimated to be $\sim 0.61,\ 0.59,\ 0.53$; within $r = 3r_{1/2} \sim 7 \ \rm kpc$.

%%%%%%%%%%%%%% subsubsection 4.3.2: half-light radius %%%%%%%%%%%%%%%
\subsubsection{CO and Optical Half-light Radii} \label{subsec:halflightradius}

%Figure 10: half-light radius comparison
\begin{figure*}[t]
\begin{center}
\includegraphics[scale = .45, bb = 0 0 819 404]{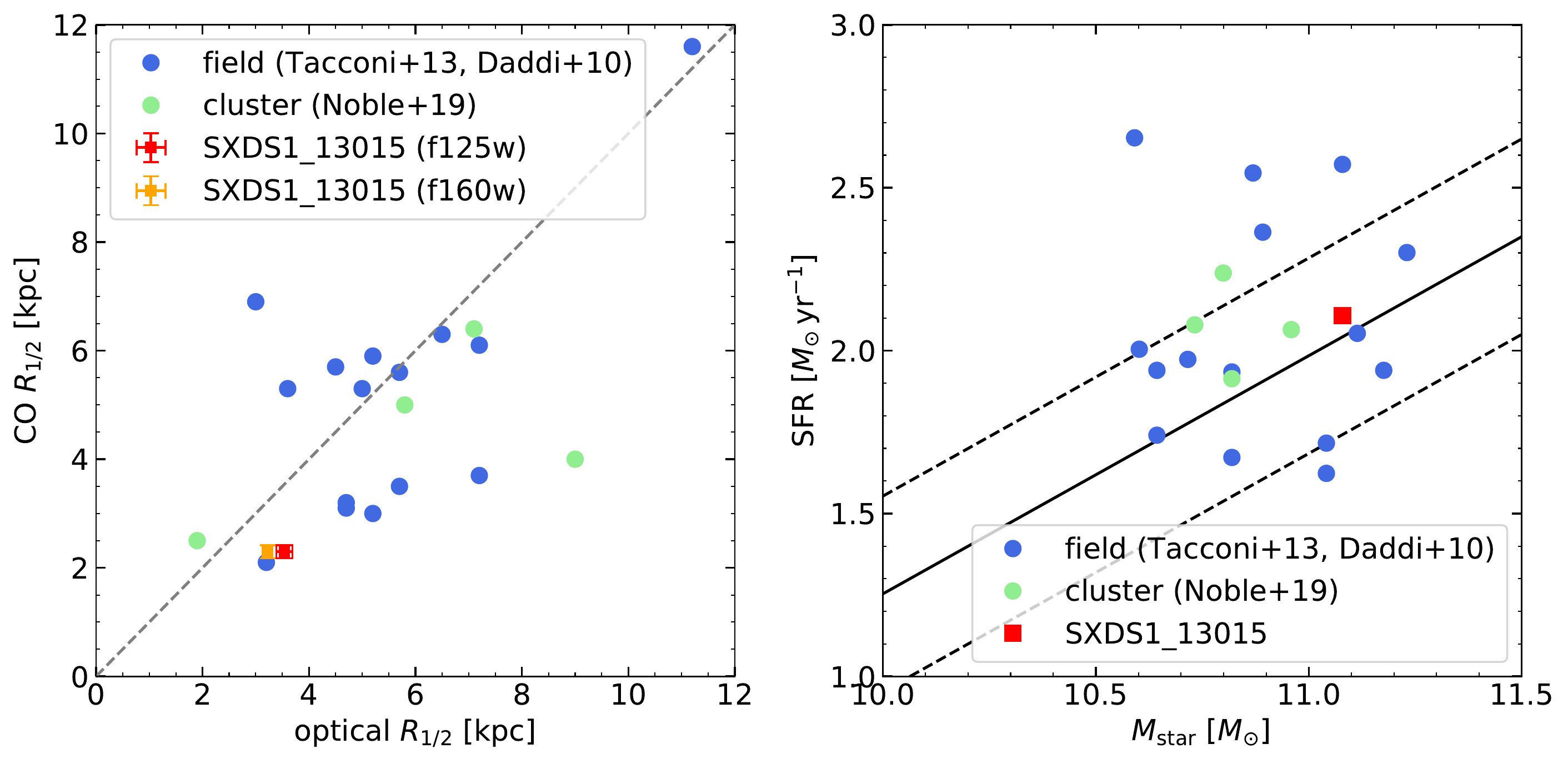}
\end{center}
\caption {\it
Left: Distribution of the galaxies in optical versus CO half-light radius plane.
SXDS1\_13015, field galaxies, and cluster galaxies are shown in red and orange squares, blue circles, and green circles, respectively.
For SXDS1\_13015, the results of the single component fittings to $J_{125}$ and $H_{160}$ images are plotted in red and orange, respectively.
The field galaxies are taken from \citet{Tacconi13} and \citet{Daddi10} (classified ``Disk(A)" in \citet{Tacconi13}).
The cluster galaxies are taken from \citet{Noble19} (with high S/N and clear velocity gradient in CO(2-1)).
The dashed line shows optical equals CO.
Right: stellar mass versus SFR.
SXDS1\_13015, field galaxies, and cluster galaxies are shown in red, blue, and green, respectively.
The main sequence at $z = 1.5$ \citep{Speagle14} is shown with solid line together with the scatter (dashed lines).
The difference of adopted IMF is corrected with the factor by \citet{Speagle14}.
}
\label{fig:radius_comparison}
\end{figure*}

Figure \ref{fig:radius_comparison} (left) shows the location of SXDS1\_13015 in optical versus CO half-light radius plane compared to other field and cluster galaxies at $z \sim 1 - 1.6$.
The field sample includes 12 main sequence galaxies at $z \sim 1 - 1.5$ by \citet{Tacconi13} and 3 sBzK galaxies by \citet{Daddi10}.
All galaxies of the field sample show both disk-like morphologies in HST images and clear CO velocity gradients (classified ``Disk(A)" by \citet{Tacconi13}).
The CO and optical half-light radii were estimated by fitting $n = 1$ S\'{e}rsic profile to image planes or circular Gaussians to {\it uv} planes, and by fitting single S\'{e}rsic profile to HST-WFC3 $I_{814}$ band data using GALFIT, respectively.
The typical uncertainties of both CO and optical half-light radius is $\sim 25\%$.
The cluster sample consists of 4 cluster galaxies at $z \sim 1.6$ by \citet{Noble19} (J0225-371, J0225-460, J0225-281 and J0225-541); they are detected with high S/Ns and show clear velocity gradients in CO(2-1).
The CO and optical half-light radii were estimated by fitting elliptical Gaussians to the CO(2-1) images and single S\'{e}rsic profile to $H_{160}$ band data using GALFIT, respectively.
The typical uncertainty of CO half-light radius is $\sim 10\%$.
As shown in Figure \ref{fig:radius_comparison} (right), the field and cluster sample galaxies are mostly main sequence galaxies at similar redshift and with similar stellar mass to SXDS1\_13015.

Studies in rest {\it V} band of galaxies at similar redshift to SXDS1\_13015 show that SFGs with similar stellar mass show half-light radius of a few to a few tens kpc and S\'{e}rsic index of $n \sim 1.5 - 4$ \citep[e.g.,][]{Wuyts11,vanderWel14,Shibuya15,Mowla19}.
The $n$ and half-light radius of SXDS1\_13015 derived from the single component fitting are within the range of those obtained in other main sequence galaxies.
The CO and optical half-light radii of SXDS1\_13015 are, however, relatively small as compared with other main sequence galaxies, but the CO-to-optical half-light radius ratio of SXDS1\_13015 ($R_{1/2,\rm CO}/R_{1/2,\rm F125W} = 0.65 \pm 0.04$, $R_{1/2,\rm CO}/R_{1/2,\rm F160W} = 0.71 \pm 0.04$) is typical among these samples.

%%%%%%%%%%%%%%%%%%%%%%%%%%%%%%%%%%%%%%%%%%%%%%%%%%%%%%%%%%%%%%%%%%%%
%%%%%%%%%%%%%%%%%%%%%%%% section 6: summary %%%%%%%%%%%%%%%%%%%%%%%%
%%%%%%%%%%%%%%%%%%%%%%%%%%%%%%%%%%%%%%%%%%%%%%%%%%%%%%%%%%%%%%%%%%%%
\section{Summary} \label{sec:summary}
We presented the results of sub-arcsecond ALMA ${\rm ^{12}}$CO(2-1) and ${\rm ^{12}}$CO(5-4) observations toward a massive main sequence galaxy at $z = 1.45$ (SXDS1\_13015).
These observations enabled us to study molecular gas properties, its distribution in the galaxy, and CO(5-4)/CO(2-1) flux ratio in a UV clump detected in the rest-frame UV HST image.
By fitting a rotating gas disk model to the CO(2-1) data, we derived molecular gas distribution and kinematics.
Combining with the HST images, we compared the properties of molecular gas and stellar distributions.
The results are as follows:

i) CO(2-1) and CO(5-4) emission lines are clearly detected from the galaxy (Figure \ref{fig:GalCO}) and the symmetric and double-peak line profile of CO(2-1) implies the presence of the molecular gas disk in the galaxy.
$R_{52}$ of the galaxy is $1.1 \pm 0.2$, which suggests the CO excitation ladder of the galaxy is similar to that of the Milky Way rather than sBzK galaxies at $z \sim 1.5$.
The molecular gas mass of the galaxy is $1.45 \times 10^{11} \ M_{\odot}$, adopting $R_{21} = 1$ (the Milky Way value) and metallicity dependent CO-to-$\rm H_2$ conversion factor.
The gas mass fraction and depletion time are $0.55 \pm 0.08$ and $1.1 \pm 0.3 \ \rm Gyr$, respectively.

ii) In the $R_{52}$ map (Figure \ref{fig:COexcitationMap}), the peak value of $R_{52}$ is $\sim 2.2$, comparable to that of the average of sBzK galaxies at $z \sim 1.5$.
$R_{52}$ peaks at the position of the UV clump.
These results are similar to a result of a hydrodynamic simulation of a clumpy galaxy \citep{Bournaud15}.
In the UV clump, the gas density and/or temperature are/is higher than those in the other galactic regions.

iii) By using $\rm GalPaK^{3D}$, the molecular gas distribution of the galaxy traced by CO(2-1) is well represented by the rotating disk model (Figure \ref{fig:Galpak}).
The half-light radius of the modeled gas disk is $2.3 \ \rm kpc$.

iv) By comparing the molecular gas distribution and stellar distribution derived by fitting S\'{e}rsic models to the rest optical HST images, the molecular gas is more concentrated in the galactic center (Figure \ref{fig:intrinsicprofile}).
The gas mass fraction decreases with increasing galactocentric radius from $\sim 0.6$ at the central region to $\sim 0.2$ at $\sim 3 \times r_{1/2}$, suggesting that the galaxy is forming its bulge.

\acknowledgments

We thank the anonymous referee for useful comments and suggestions, which improve the paper.
We are grateful to K. Nakanishi, F. Egusa, K. Saigo,  and the staff at the ALMA Regional Center for their help in data reduction.
We also thank K. Tadaki for his critical comment on our earlier study with this galaxy.
F.M. is supported by Research Fellowship for Young Scientists from the Japan Society of the Promotion of Science (JSPS).
K.O. is supported by JSPS KAKENHI Grant Number JP19K03928.
This paper makes use of the following ALMA data: ADS/JAO.ALMA\#2015.1.01129.S. and \#2011.0.00648.S.
ALMA is a partnership of ESO (representing its member states), NSF (USA) and NINS (Japan), together with NRC (Canada), MOST and ASIAA (Taiwan), and KASI (Republic of Korea), in cooperation with the Republic of Chile. The Joint ALMA Observatory is operated by ESO, AUI/NRAO and NAOJ.

\software{GALFIT \citep{Peng02}, GalPaK3D \citep{Bouche15}, CASA \citep[v4.7.2;][]{McMullin07}}

%%%%%%%%%%%%%%%%%%%%%%%%%%%%%%%%%%%%%%%%%%%%%%%%%%%%%%%%%%%%%%%%%%%%
%%%%%%%%%%%%%%%%%%%%%%%%%%% bibliography %%%%%%%%%%%%%%%%%%%%%%%%%%%
%%%%%%%%%%%%%%%%%%%%%%%%%%%%%%%%%%%%%%%%%%%%%%%%%%%%%%%%%%%%%%%%%%%%
\bibliography{ref_13015}{}
\bibliographystyle{aasjournal}

%%%%%%%%%%%%%%%%%%%%%%%%%%%%%%%%%%%%%%%%%%%%%%%%%%%%%%%%%%%%%%%%%%%%
%%%%%%%%%%%%%%%%%%%%%%%%%%%%%%%%%%%%%%%%%%%%%%%%%%%%%%%%%%%%%%%%%%%%
%%%%%%%%%%%%%%%%%%%%%%%% end of the document %%%%%%%%%%%%%%%%%%%%%%%
%%%%%%%%%%%%%%%%%%%%%%%%%%%%%%%%%%%%%%%%%%%%%%%%%%%%%%%%%%%%%%%%%%%%
%%%%%%%%%%%%%%%%%%%%%%%%%%%%%%%%%%%%%%%%%%%%%%%%%%%%%%%%%%%%%%%%%%%%
\end{document}